\documentclass[twoside,fleqn]{article}
\usepackage{espcrc2}
\usepackage[dvips]{graphicx}
\def\be{\begin{equation}}
\def\ee{\end{equation}}
\def\bea{\begin{eqnarray}}
\def\eea{\end{eqnarray}}
\def\sbs{$SB\chi S$}
\newcommand{\psim}{{{}<\atop{}^\sim}}

\newcommand{\AmS}{{\protect\the\textfont2
  A\kern-.1667em\lower.5ex\hbox{M}\kern-.125emS}}

\title{Theoretical and Phenomenological Aspects of $G\chi PT$\thanks{
IPNO/TH 97-29}\thanks{
Invited talk at the workshop QCD97, Montpellier, France, July  1997}}

\author{Jan Stern\address{Division de Physique Th\'eorique,
              Institut de Physique Nucl\'eaire, \\
              91406 Orsay, France}%
        \thanks{Laboratoire des Universit\'es Paris XI et Paris VI associ\'e au
CNRS.}}

\begin{document}

\begin{abstract}
Consequences of the alternative mechanism of spontaneous breaking of chiral
symmetry (\sbs) without a formation of a large quark antiquark condensate is
reviewed. Emphasis is put on the resulting picture of light quark masses.
\end{abstract}

\maketitle

\section{THEORETICAL ASPECTS OF $SB\chi S$ IN QCD}

The spontaneous breaking of chiral symmetry $SU_L(N_f)\times
SU_R(N_f)\times U_V(1)$,  (associated  with $N_f$  massless flavours), down to
the diagonal subgroup $U_V(N_f)$ is one of our theoretical certitudes in QCD
\cite{H,V}. It is equivalent to the statement that in the chiral limit $m\to 0$,
the left-right correlator
\bea
F^2\delta^{ij}=\lim_{m\to 0}i\eta^{\mu\nu}
\int dx\
<\Omega\vert TJ^{i}_{\mu L}(x)\nonumber\\
J^j_{\nu R}(0)\vert\Omega >\ne 0,
\label{F2}
\eea
where $J_{\mu L},\ (J_{\mu R})$ are Noether currents generating left (right)
chiral rotations, remains non-zero. (F is nothing but the pion decay constant
$F_\pi=92.4$ MeV at $m=0$). Indeed, Eq. (\ref{F2}) encompasses both essential
features of $SB\chi S$ : {\bf i)} The asymmetry of the vacuum and, {\bf ii)}
the existence of $N^2_f-1$ massless Goldstone bosons coupled to the Noether
currents with the strength F. In other words, F is an {\bf order parameter},
whose non-vanishing is not only a sufficient but also a necessary condition of
\sbs. There are, of course, many other order parameters, such as the $q\bar q$
condensate $(N_f=2,\ m_u=m_d=m)$
\be
\lim_{m\to 0}
<\Omega\vert\bar u u\vert\Omega>
=\lim_{m\to 0}<\Omega\vert\bar d d\vert\Omega>
=-F^2B,
\label{lim}
\end{equation}
playing a different role than the pion decay constant F : $B\ne 0$ is {\bf not}
a necessary consequence of \sbs, rather, the magnitude of B reflects the nature
of the chiral order in the QCD vacuum. Quark condensate is analogous to the
spontaneous magnetization $\vec m$ of spin systems, reflecting the type of the
magnetic order in the ground state : $\vec m\ne 0$ for a ferromagnet, but
$\vec m=0$ for an antiferromagnet.

The non-equivalence of quark condensation and of \sbs\ may be illustrated
treating the Euclidean QCD as a disordered system\cite{V}. Indeed, the
mechanism of \sbs\ can be analyzed in terms of basic characteristics of lowest
eigenstates of (Euclidean) Dirac operator in a random gluonic background :
density and localization. A $q\bar q$ condensate is formed, provided the
density of such states is large enough : the average number of eigenvalues in a
given small neighbourhood of zero should be proportional to the volume
V\cite{B}. On the other hand, $F_\pi$ - Eq. (\ref{F2}) - remains non zero, i.e.
\sbs\ takes place and Goldstone bosons are formed under somewhat weaker
conditions. $F\ne 0$ can actually be realized {\bf either} with a {\bf large}
density of {\bf localized} states, {\bf or} with a {\bf lower} density (number
of small eigen values $\sim\sqrt{V}$) of {\bf delocalized} states. In the first
case quarks do condense $(B\ne 0)$, whereas in the second case $B\to 0$ in the
thermodynamic limit $V\to \infty$\cite{S}.

The question of the size of the quark condensate is of theoretical nature :
there is no known way how to decide a priori which of the two mechanisms
summarized above takes actually place and, even if $B\ne 0$, how close are we
to the pure $B=0$ scenario. In practice, one has to decide whether $B$ is as
large as $\sim$ 2 GeV, as suggested by lattice simulations\cite{G}, or as small
as $B\sim$ 100 MeV as it results from a recent attempt to calculate directly in
the continuum within the framework of variationally improved perturbation
theory\cite{N}. The decision will likely come from new low-energy high
precision experiments\cite{A,Ba}.

\section{GENERALIZED CHIRAL PERTURBATION THEORY}

In Nature, the $\overline{MS}$ running quark masses $m_u,m_d,m_s$
(renormalized at the QCD scale $\nu$ = 1 GeV) appear small compared to the
typical mass-scale of the bound states $(\rho, N,\cdots),\ \Lambda_H\simeq$ 1
GeV. Low energy observables can be expanded in powers of $m/\Lambda_H$ and in
powers of external momenta $p/\Lambda_H$. The standard version of $\chi PT$
assumes, in addition,  that the deviation $\Delta$ from the GOR relation\cite{GOR}
\be
M^2_\pi=2mB+\Delta
\label{M2}
\end{equation}
can be treated as a small perturbation : $S\chi PT$\cite{GL} is a simultaneous
expansion in powers of $m_q/\Lambda_H$ {\bf and} of
$\epsilon=\Delta/ M^2_\pi$. If B vanishes, $\epsilon=1$ and the latter
expansion breaks down. For nonzero but sufficiently small B, this expansion
could still break down, unless quark masses are mathematically small. $G\chi
PT$\cite{F,K} still expands in powers of $m/\Lambda_H$ avoiding however any
expansaion in powers of $\epsilon=\Delta/M^2_\pi$. Technically, this is
achieved by a different chiral power counting\cite{F} :
\be
\begin{array}{lll}
S\chi PT : & m_q=O(p^2)\ , & B=O(1)\\
G\chi PT : & m_q=O(p)\ , &   B=O(p)\ .
\end{array}
\label{mat}
\end{equation}
In $G\chi PT$, the first term in the expansion (\ref{M2}) of $M^2_\pi$ is
suppressed, and this is why $2mB$ and $\Delta$ could be of a comparable size.

\section{EXPANSION OF $M^2_\pi$}

It is instructive to consider in more details the expansion $(2m=m_u+m_d)$
\bea
{1\over F^2}F^2_\pi M^2_\pi=2mB +\hspace{3cm}\nonumber \\
+ 4m^2
\left\{
A(\mu)-{3B^2\over 32\pi^2F^2}\ln{M^2_\pi\over\mu^2}
\right\}
+O(m^3)\hspace{0.5cm}
\label{1/F2}
\eea
which, as it stands, is independent of the chiral counting (\ref{mat}). The
point we would like to make is that the constant $A(\mu)$ (for
$\mu\sim\Lambda_H$) is insensitive to the Goldstone-boson sector of the theory,
in particular, to the size of B. This can be seen from the definition of the
constant $A(\mu)$ in terms of the two point fucntion of scalar-isoscalar and
pseudoscalar-isovector quark densities  $S^0(x)$ and $P^{i}(x)$. For
$m_u=m_d=0$, one has \cite{GL}
\bea
{i\over F^2}\int dx e^{ipx}
<0\vert T\{\delta^{ij}S^0(x)S^0(0)-\nonumber \\
-P^{i}(x)P^j(0)\}\vert 0>=\nonumber \\
=\delta^{ij}
\left\{
{B^2\over p^2}+{3\over 32\pi^2}\ {B^2\over F^2}
\left(
\ln{\mu^2\over -p^2}+1
\right)+A(\mu)\right.\nonumber \\
\left.+O(p^2)
\right\}.
\label{i/F2}
\eea
The first term on the r.h.s. is the GB-pole, the second term results from the
GB-loop, whereas $A(\mu)$ collects contributions from exchanges of
non-Goldstone particles : $\sigma, \pi'\cdots$It is hard to believe that the
latter would be significantly affected by the mechanism of \sbs. Comparing Eqs.
(\ref{1/F2}) and (\ref{i/F2}) with the corresponding $S\chi PT$
expressions\cite{GL}, one finds
\be
A(\mu)={2B^2\over F^2}
\left[
l^r_3(\mu)+l^r_4(\mu)
\right].
\label{Amu}
\end{equation}
This allows one to estimate $A(\mu)$ in the standard case of a large condensate
: Taking e.g. $B\simeq$ 1.6 GeV (corresponding to $m\simeq$ 6.1 MeV) and using
the standard values \cite{GL}, $l^r_3(M_\rho)=0.8\times 10^{-3},\
l^r_4(M_\rho)=5.6\times 10^{-3}$, one finds
\be
A(M_\rho)\simeq 4\ .
\label{Mro}
\end{equation}
This value should not be considerably modified in $G\chi PT$. Consequently, if
B drops out, the low energy constant $l_3+l_4$ of $S\chi PT$ should blow up,
signalizing the inadequacy of the standard expansion, rather than a pathology.
If B is small enough to justify the $G\chi PT$ counting (\ref{mat}), the
expansion (\ref{1/F2}) can be rewritten as
\be
M^2_\pi=2mB+4m^2A+O(p^3)\ .
\label{op3}
\end{equation}
The first two terms are now of the same order $O(p^2)$. The NLO $O(p^3)$
correction includes the factor ${F^2_\pi/F^2}-1$ as well as a part of
$O(m^3)$ contributions in Eq. (\ref{1/F2}). Finally, the chiral logarithms
proportional to $m^2B^2$ are now suppressed : they are relegated to the order
$O(p^4)$. The linear and quadratic terms in Eq. (\ref{op3}) are of comparable
size for
\be
m\sim m_0={B\over 2A}\ .
\label{mmo}
\end{equation}
For $B\sim$ 100 MeV, $A\sim 4$, $m_0$ can, indeed, be so small to make
problematic any expansion in powers of $m/m_0$, in particular, the $S\chi PT$.

\section{LOW-ENERGY $\pi\pi$ SCATTERING}

The low-energy $\pi\pi$ scattering amplitude is particularly sensitive to the
size of the $\bar q q$ condensate\cite{F,KM}.
It's $S\chi PT$ expansion\cite{W} makes appear the low energy constant $l_3$
and not just the combination $B^2l_3$ as in Eqs. (\ref{1/F2}) and (\ref{i/F2}).
It is instructive to look at the low energy parameter $\alpha$
(defined including the $O(p^6)$ accuracy in \cite{KM}), which is
closely related to the scattering amplitude $A(s,t,u)$
at the symmetric
point $s=t=u={4\over 3}M^2_\pi$:
\be
A
\left(
{4\over 3}M^2_\pi,{4\over 3}M^2_\pi,{4\over 3}M^2_\pi
\right)
={\alpha\over 3F^2_\pi}M^2_\pi\ + \cdots
\label{A43}
\end{equation}
The parameter $\alpha$ turns out to be more sensitive to the magnitude of B
than, say, the S-wave scattering lengths : At the leading order of $G\chi PT$,
$\alpha = 1$ in the case of a large condensate, whereas
$\alpha\simeq 4$ for $B = 0$. In addition, the parameter $\alpha$
receives rather moderate corrections from higher orders of $\chi PT$. Nowaday,
the $\pi\pi$ scattering amplitude is analyzed up to and including two-loop
accuracy both in the standard\cite {Bi} and in the generalized\cite{KM}
frameworks. Within the $S\chi PT$ one gets for instance
\bea
\nonumber
\label{al}
\alpha=&\hspace{-0.3cm} 1+& \\
&~& \hspace{-1.5cm}
\left(
6l^r_3+2l^r_4-{1\over 32\pi^2}\ln{M^2_\pi\over\mu^2}-{1\over 32\pi^2}
\right)
{M^2_\pi\over F^2_\pi}+\cdots
\eea
leading to the prediction\cite{Gi}
\be
\alpha=1+0.06+0.01+\cdots=1.07\pm 0.007\ .
\label{chi}
\end{equation}
(Here, the contributions of successive orders $O(p^2),\ O(p^4)$ and $O(p^6)$
are explicitely shown). This has to be compared with the experimental
value\cite{KM}
\be
\alpha_{exp}=2.16\pm 0.86
\label{exp}
\end{equation}
inferred from the Geneva-Saclay $K_{l4}$\cite{R} and CERN-M\"unich $\pi
N\to\pi\pi N$\cite{Hy} data, using the technique of Roy dispersion relations.
Notice that the result (\ref{exp}) implies for the $q\bar q$ condensate
$2mB=(0.6\pm 0.4)M^2_\pi$. Results of new high precision experiments, such as
the $\pi^+\pi^-$ - atom lifetime measurement at CERN\cite{A} and
the new $K^{+}_{l4}$ experiments \cite{Ba} at BNL and
Frascati are awaited with an obvious interest.

\section{THE STRANGE QUARK MASS}

Treating also the s-quark as light, the $G\chi PT$ formula (\ref{op3}) can be
extended to remaining unmixed Goldstone bosons :
\be
\begin{array}{l}
M^2_\pi=2mB_0+4m^2A_0+O(p^3)\\
M^2_K=(m_s+m)B_0+(m_s+m)^2A_0+O(p^3)\ .
\end{array}
\label{mK}
\end{equation}
The constants $B_0$ and $A_0$ differ from $B$ and $A$ introduced before by
Zweig
rule violating effects, which are expected to be small. Eq. (\ref{mK}) implies
(the dots stand for higher order corrections)
\bea
{2 mB_0\over M^2_\pi} & = & {(r-r_1)(r+r_1+2)\over r^2-1}
+\cdots
\label{2m}\\
r & = & r_2-2A_0
\left(
{m_s\over M_\pi}
\right)^2
+\cdots\ ,
\label{rr2}
\eea
where $r$ is the quark mass ratio $r={m_s / m}$ and
\be
r_n=2
\left(
{M_K\over M_\pi}
\right)^n-1\ ,\ n=1,2\ ,
\label{rn}
\end{equation}
i.e. $r_1\simeq 6.3$ and $r_2\simeq 25.9$. Notice that, according to Eq.
(\ref{2m}), the stability of the vacuum condition $B_0\geq 0$, requires $r\geq
r_1\simeq 6.3+\cdots$. It is instructive to look at Eq. (\ref{rr2}) in the
light of the new {\bf experimental} information on $m_s$, obtained by the
ALEPH-collaboration from the precise measurement of the inclusive rate of the
$\tau$-decays $\tau\to \nu_\tau+X(S=1)$. The preliminary result reads\cite{CH}
\be
m_s\ \mbox{(1 GeV)}=(235\ ^{+35}_{-42})\mbox{MeV}
\label{ms}
\end{equation}
and it is somewhat higher than expected on the basis of sum rule\cite{J} and
lattice\cite{Lat} estimates.

A higher value of $m_s$ is welcome for the $G\chi PT$ scenario : It {\bf
lowers} the values of the quark mass ratio $r$ and of the condensate parameter
$B_0$. To illustrate this point by a crude numerical estimate, let us take as
the input (at the QCD scale $\nu=$ 1 GeV), $m_s=$ 200 MeV and $A_0\simeq
A\simeq 4$ according to Eq. (\ref{Mro}). Eq. (\ref{rr2}) then yields $r\simeq
 8.3$, implying in turn, through Eq. (\ref{2m}), $2 mB_0\simeq 0.5\
M^2_\pi$, in agreement with the available information from $\pi\pi$-scattering.
The condensate should then be $B_0\simeq 190$ MeV and $m$ (1 GeV)
$\simeq 24$ MeV. Let us stress once again that these numbers constitute a crude
estimates with neither error bars nor higher order effects included. A more
detailed analysis will be given elsewhere\cite{FKS}.

\section{$m_u+m_d$ AND QCD SUM RULES}

We finally discuss the constraints imposed on the value of $m={1\over
2}(m_u+m_d)$ by the QCD sum rules involving the two point function
\bea
\Psi_5(q^2) & = & i
\int dx\ e^{iqx}\times\nonumber\\
&&\times<\Omega\vert TD_5(x)D_5^+(0)\vert\Omega>\ ,
\label{psi5}
\eea
where $D_5$ stands for the divergence of the axial current
\be
D_5\equiv \partial^\mu\bar u\gamma_\mu\gamma_5 d=2m\bar u i\gamma_5 d\ .
\label{D5}
\ee
Evaluation of these sum rules needs information about the imaginary part
\be
{1\over\pi}\mbox{Im}\Psi_5(t)=2F^2_\pi M^4_\pi\delta(t-M^2_\pi)+\rho(t)\ .
\label{Im}
\ee
The spectral function $\rho(t)$ collects all the continuum contributions of
$3\pi,\ K\bar K\pi,\ 5\pi\cdots$ intermediate states. It is in principle
measurable in high statistics $\tau$-decays \cite{St} but so far, both its shape and
its {\bf normalization} are unknown. $\rho(t)$ is a product of a genuine
massless QCD correlation function (similar to the one in Eq. (\ref{i/F2})) and
of $m^2$. Since, in QCD, m is a free parameter, there is no way how to deduce
the normalization of $\rho$ from the theory, without an independent
experimental input.

Two devices have been used in the literature to circumvent this difficulty. The
first one attempts to use the $\chi PT$ expression of $\rho(t)$ near threshold
to guess its normalization in the resonance region \cite{D}. It turns out that
this $\chi PT$-based normalization is surprisingly sensitive to the size of the
$q\bar q$ condensate. At {\bf threshold}, one has up to the leading order
\cite{St,K}
\be
\rho\vert_{B=0}=13.5\times\rho\vert_{S\chi PT}\ .
\label{135}
\ee
The $G\chi PT$ knows that a substantially larger value of the quark mass m is
needed for its internal consistency.

The second attempt to constrain the normalization of $\rho$ makes use of
QCD-hadron duality \cite{P} : It requires \cite{D} that the ratio
\be
R_{Had}(s)={3\over 2s}
{\int_0^sdt\ t\rho(t)\over
2F^2_\pi M^4_\pi+
\int_0^sdt\rho(t)}
\label{Rhad}
\ee
coincides in a suitable range of $s$ with the corresponding QCD expression
$R_{QCD}(s)$ which is given by OPE and is almost independent of $m$. For small
spectral densities, $R_{Had}$ is, indeed, rather sensistive to the
normalization of $\rho$. However, to the extent that the continuum contribution
in the denominator  of (\ref{Rhad}) dominates over the pion contribution, Eq.
$R_{Had}=R_{QCD}$ becomes independent of the normalization of $\rho$,
constraining merely its shape. This is what is {\bf expected} to happen as a
consequence of the $G\chi PT$ power counting (\ref{mat}) : $\int dt\rho(t)$
counts as $O(p^2)$, whereas the pion contribution is $O(p^4)$. It is actually
not difficult to construct a model of the shape of $\rho(t)$, for which the
duality criterion would be satisfied \cite{FKS} {\bf without including the
contribution of the pion pole}. After all, in the scalar channel there is no
pion pole and the corresponding $R_{QCD}(s)$ are about the same in both
pseudoscalar and scalar channels.

Even if the normalization of the spectral function remains unknown, the QCD sum
rules can still be combined with the Ward identity
\be
\Psi_5(0)=-2m<\Omega\vert\bar uu+\bar dd\vert\Omega>
\label{-2m}
\ee
to investigate the variation of $\Psi_5(0)$ with the quark mass
$m$, {\bf
keeping fixed $F^2_\pi M^2_\pi$} at its experimental value. Notice that
(\ref{-2m}) does not directly involve the condensate (\ref{lim}) which is
defined in the chiral limit. For small $m$, however,
\be
{1\over F^2}\Psi_5(0)=4mB+2m^2C+\cdots\ ,
\label{4B}
\ee
where $C$ is a high energy counterterm which depends on the way one
renormalises $\Psi_5(0)$. We now consider \cite{FKS} the Laplace transform sum
rule for $\Psi_5(-Q^2)/Q^2$ making appear $\Psi_5(0)$, together with its
derivatives. We use a simple model for the shape of the spectral function :
\bea
\rho(t)=[m(\mbox{1 GeV})]^2
\left\{
g^2[\delta(t-M^2_1)+\right.\nonumber\\
\left.
+x\delta(t-M^2_2)]
+\gamma_{as}(t)
\theta (t-t_{0}) \right\}
\label{rot}
\eea
where $M_1$ = 1.30 GeV, $M_2$ = 1.77 GeV, $x\simeq 1$ and $\gamma_{as}(t)$ is
given by QCD asymptotics. Eliminating the unknown constant $g^2$ one obtains
for each value of $m$ a prediction for the condensate ratio $\Psi_5(0)/2F^2_\pi
M^2_\pi$, which in $S\chi PT$ is expected to be close to 1. The result is
displayed in Fig. 1.
\begin{figure}[h]
\begin{center}
\includegraphics*[scale=0.33,angle=-90]{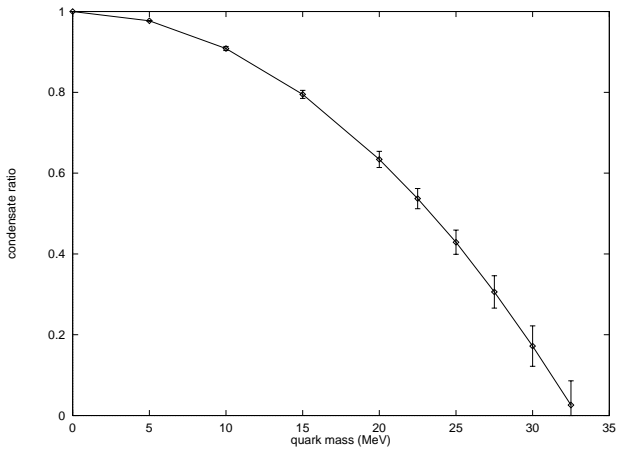}
\caption{\sl The condensate ratio $\Psi_5(0)/2F^2_\pi M^2_\pi$ as a
function of $m$(1GeV) from Borel transform sum rules for $\Psi_5(q^2)$.
}
\end{center}
\end{figure}
The indicated errors reflect the variation of individual
output values with the Borel variable $u$. The shape of the curve on Fig. 1 can
be understood within $G\chi PT$ : Using the expansion (\ref{1/F2}) and Eq.
(\ref{4B}), one gets
\bea
-{m\over F^2_\pi M^2_\pi}
&&<\Omega\vert\bar uu+\bar dd\vert\Omega>  =  \nonumberÊ\\
 &&=  1-4{m^2\over M^2_\pi}(A-{1\over 4}C)+\cdots
\label{1-4}
\eea
which reproduces the curve in the whole range of $m\psim 35$ MeV, provided one
takes
\be
A-{1\over 4}C\simeq 4\div 5.4
\label{AC}
\ee
(independent of $m$). It is instructive to compare this result with our
estimate (\ref{Mro}).

\section{SUMMARY}

The leading order $G\chi PT$ expression of $M^2_\pi$ and $M^2_K$ involves {\bf
i)} the condensate parameter $B_0$, {\bf ii)} the quadratic slope parameter
$A_0$, {\bf iii)} $m={1\over 2}(m_u+m_d)$ and {\bf iv)} $m_s$. Consequently, in
order to get a complete picture of quark masses and condensates, , it is
sufficient to pin down any two of these 4 parameters independently of any
(hidden) prejudices about the size of the $q\bar q$ condensate. I have
suggested to use recent experimental determination of $m_s$ by the ALEPH
collaboration, together with an estimate of $A_0$. I have argued that $A_0$
should be roughly the same both in $S\chi PT$ and in $G\chi PT$, rather
insensitive to the $q\bar q$-condensation. The crude estimate $A_0\simeq 4$
(with errors hard to determine) then leads to a rather small values of
$r=m_s/m\simeq 8.3$ and of the condensate parameter $B\sim$ 190 MeV. $2mB$ then
represents half of $M^2_\pi$, in agreement with existing $\pi-\pi$ scattering
data. The corresponding large value of $m\simeq 24$ MeV has been argued to be
compatible with relevant QCD sum rules and with QCD-hadron duality, provided
the hadronic spectral function is indeed larger than usually expected. This can
be tested measuring azimuthal asymmetries in $\tau\to 3\pi+\nu_\tau$ \cite{St}.

The preceeding discussion certainly lacks a quantitative precision and it
should be merely viewed as a plausibility argument illustrating the coherence
of the small condensate scenario. The precise quantitative test is awaited in
low-energy $\pi^+\pi^-$ scattering.

I am indebted to Norman Fuchs, Marc Knecht and Bachir Moussallam for
discussions and help in collecting materials for this talk.

\end{document}